\documentclass{jpsj-suppl}
\usepackage{txfonts,color} 

\title{Valence Fluctuations in the Extended Periodic Anderson Model at
Finite Temperatures}

\author{Yuhei \textsc{Kojima} and Akihisa \textsc{Koga}}

\inst{Department of Physics, Tokyo Institute of Technology, 
Tokyo 152-8551, Japan}

\email{koga@phys.titech.ac.jp}

\recdate{July 14, 2013}

\abst{
We study valence fluctuation at finite temperatures 
in the periodic Anderson model with 
the Coulomb interaction between $f$ and conduction electons,
combining dynamical mean-field theory with the non-crossing approximation. 
It is found that a large Coulomb repulsion induces 
a first-order valence transition. 
The finite temperature phase diagram is determined.
}

\kword{valence transition, dynamical mean-field theory, 
non-crossing approximation}

\begin{document}
\maketitle

\section{Introduction}
Since the discovery of heavy-fermion materials with rare-earth
or actinide elements, this class of strongly correlated electron systems 
has attracted considerable attention. 
Among them, valence fluctuation phenomena and valence transitions 
in some Ce- and Yb-based compounds 
are interesting topics in this field~\cite{Ce,Yb1,Yb2}.
It has recently been suggested that 
critical valence fluctuations play an important role 
in understanding non-fermi liquid behavior and 
superconductivity in a certain compounds~\cite{Onishi,Miyake},
which stimulates further theoretical 
investigations~\cite{Watanabe,Saiga,Sugibayashi}.
Valence fluctuations have theoretically been discussed by means of
the periodic Anderson model with 
the interaction between the conduction and $f$-electrons. 
It has been clarified that
the interaction plays
an important role in stabilizing the valence transition,
where the Kondo state competes with the mixed-valence state.
On the other hand, it may not be so clear 
how stable a valence transition is against thermal fluctuations.
To clarify this, we consider the extended periodic Anderson model.
We make use of dynamical mean-field theory (DMFT)
\cite{Metzner,Muller,Pruschke95,Georges96,Kotliar04} and 
the non-crossing approximation (NCA)~\cite{Kuramoto,Eckstein}, 
which allows us  
to discuss how valence transitions are realized at finite temperatures.

\section{Model and Method}

The system should be described by the following
extended periodic Anderson Hamiltonian 
\cite{Schork,Sato,KogaPAM,KogaPAM2,Yoshida} as,
\begin{eqnarray}
H &=& \sum_{(ij)\sigma} t_{ij} c_{i\sigma}^\dagger c_{j\sigma} 
+ \epsilon_f \sum_{i\sigma} f_{i\sigma}^\dagger f_{i\sigma} 
+ V\sum_{i\sigma} (c_{i\sigma}^\dagger f_{i\sigma} + H.c.)\nonumber\\
   &+&  U\sum_i n^{f}_{i \uparrow} n^f_{i \downarrow} + U_{cf} \sum_{i\sigma \sigma'} n^c_{i \sigma} n^f_{i \sigma'},
\label{Hami}
\end{eqnarray}
where $f_{i\sigma} (c_{i\sigma})$ annihilates an $f$-electron 
(conduction electron) with spin $\sigma(=\uparrow, \downarrow)$ 
at the $i$th site, $n^c_{i\sigma}=c_{i\sigma}^\dag c_{i\sigma}$, and 
$n^f_{i\sigma}=f_{i\sigma}^\dag f_{i\sigma}$.
Here, $t_{ij}$ represents the hopping integral for conduction electrons, 
$\epsilon_f$ is the energy level of the $f$ state, and $V$
the hybridization between the conduction and $f$ states.
$U$ is the Coulomb interaction between $f$-electrons and $U_{cf}$ is
the Coulomb interaction between conduction and $f$ electrons.

To investigate the extended periodic Anderson model eq. (\ref{Hami}),
we make use of DMFT~\cite{Metzner,Muller,Pruschke95,Georges96,Kotliar04}.
which has successfully been applied 
to various strongly correlated electron systems.
In the framework of DMFT, the lattice model is mapped to
 an effective impurity  model, 
where local particle correlations are taken into account precisely. 
The Green function  for the original lattice system is then obtained 
via self-consistent equations imposed on the impurity problem.

In DMFT,
the Green function in the lattice system is given as,

\begin{equation}
{\hat G}_{\sigma}({\bf k}, z)^{-1}=
\left(
\begin{array}{cc}
z +\mu -\epsilon_k & -V \\
 -V  & z +\mu -\epsilon_f 
\end{array}
\right)-
\left(
\begin{array}{cc}
\Sigma_{cc}(z) & \Sigma_{cf}(z) \\
\Sigma_{fc}(z)  & \Sigma_{ff}(z) 
\end{array}
\right),
\end{equation}
where $\mu$ and $\Sigma$ are the chemical potential and 
the self-energy, and 
$\epsilon_k$ is the dispersion relation for the bare conduction band.
In the following, we consider the $D$-dimensional Bethe lattice 
with the hopping integral $t=t^*/D$, which results in the 
density of state $\rho_0(x)=\frac{2}{\pi D}\sqrt{1-(x/D)^2}$.
Then the self-consistency condition is
represented as~\cite{Schork},
\begin{eqnarray}
\left[\hat{\cal G}(z)^{-1}\right]_{cc}&=&z+\mu
-\left(\frac{D}{2}\right)^2 G_{cc}(z),
\end{eqnarray}
where $\hat{\cal G}$ is the non-interacting Green function of 
the effective impurity model.

There are various numerical methods to solve the effective impurity problem.
To discuss how valence fluctuations induce transitions 
in the extended periodic Anderson model,
we use here the NCA method,
which allows us to discuss finite-temperature properties quantitatively. 
In the following, we take $D=1$ as unit of energy and use the parameters
$U=20$ and $V=0.25$. We also fix the total number of particles as 
$n=1.9$, where $n=\sum_{i\sigma}\langle n_{i\sigma}^c+n_{i\sigma}^f\rangle /N$ 
and $N$ is the total number of sites.

\section{Results}

\begin{figure}[htb]
\begin{center}
\includegraphics[width=9cm]{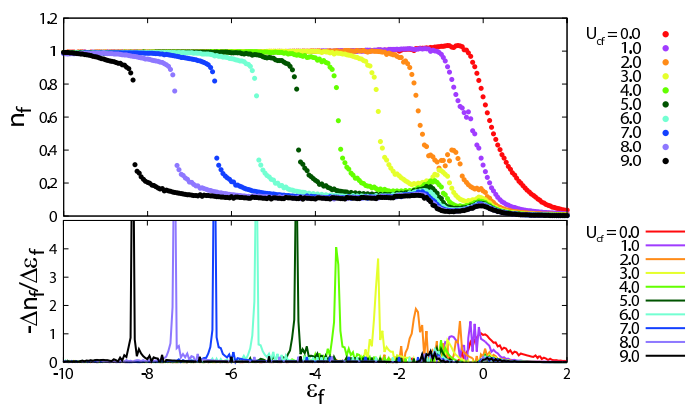}
\end{center}
\caption{(Upper panel) the $f$-electron number 
(lower panel) its derivative as a function of the $f$-level 
at the temperature $T=0.05$ when $U_{cf}=0.0, 1.0, \cdots,$ and $9.0$. 
}
\label{1}
\end{figure}

We consider valence fluctuations in the model 
when the interaction between the conduction and $f$-electons varies.
To this end, we calculate the number of $f$-electrons 
$n_f\left(=\sum_{i\sigma}\langle n_{i\sigma}^f \rangle /N\right)$ and 
its numerical derivative with respect to $\epsilon_f$,
as shown in Fig. \ref{1}.
When the $f$ level is low enough and the interaction between $f$-electrons 
is strong enough, the heavy-metallic Kondo state is realized, 
where $n_f\sim 1$.
On the other hand, as the $f$ level becomes higher, 
$n_f$ is away from the commensurate value, where the mixed-valence 
state is realized.
It is found that when $U_{cf}=0$, 
the Kondo state is continuously changed to the mixed-valence state
and the crossover occurs around $\epsilon_f\sim 0$,
where $|\Delta n_f/\Delta \epsilon_f|$ has maximum.
On the other hand, as increasing $U_{cf}$,
different behavior appears.
The valence at the $f$ level is rapidly changed around the crossover region.
Beyond the critical value of $U_{cf}$,
the jump singularity appears in the curves, as shown in Fig. \ref{1}.
This implies the existence of a first-order phase transition between
the Kondo and mixed-valence states.

\begin{figure}[htb]
\begin{center}
\includegraphics[width=8cm]{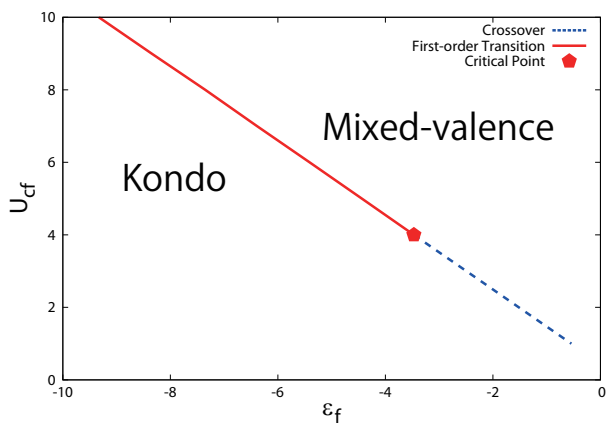}
\end{center}
\caption{The phase diagram of the extended periodic Anderson model 
at the temperature $T=0.02$. Solid line represents 
the phase boundary between the Kondo and mixed-valence regions.
Dashed line indicates the crossover between two states.
}
\label{2}
\end{figure}

By performing similar calculations, we obtain the finite
temperature phase diagram, as shown in Fig. \ref{2}. 
The Kondo state with $n_f\sim 1$ is realized in the case with lower $f$-level.
Increasing the $f$ energy level, valence fluctuations 
are enhanced.
In the case $U_{cf}>4$, the first-order valence transition occurs
from the Kondo state to the mixed-valence state,
while in the other, no sigularity appears in $n_f$ and the crossover appears. 
At the end point of the phase boundary $(U_{cf}, \epsilon_f)\sim (4, -3.7)$, 
the derivative of the particle number of $f$-electron 
diverges, where the phase transition is of second order.

\begin{figure}[htb]
\begin{center}
\includegraphics[width=8cm]{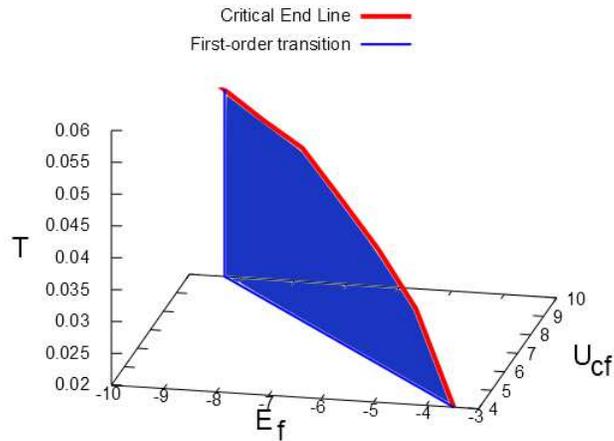}
\end{center}
\caption{The phase diagram in $T-U_{cf}-\epsilon_f$ space.
}
\label{3}
\end{figure}

Fig. \ref{3} shows the phase diagram in $T-U_{cf}-\epsilon_f$ space, 
by estimating the phase boundaries in the system at different temperatures
systematically.
At higher temperatures, the effect of the interaction becomes irrelevant and
the Kondo state with $n_f\sim 1$ becomes unstable.
Therefore, the jump singularity in the particle number of $f$-electons smears 
with increasing temperatures,
where the critical end point is shifted up,
as shown in Fig. \ref{3}.
An important point is that the first-order phase boudary between
Kondo and mixed-valence regions is little affected 
by the thermal fluctuation.
This implies that at least, in the framework of DMFT with NCA,
the valence transition is not induced by lowering temperatures but
is induced by changing local parameters $U_{cf}$ and $\epsilon_f$.
On the other hand, at low temperatures, the NCA method may not be appropriate
to discuss low-energy properties in the system quantitatively.
Nevertheless, our results suggests the existence of the quantum critical point 
at $T=0$,
which is consistent with the previous works~\cite{Watanabe,Saiga}.
Therefore, we have discussed the valence transitions 
in the extended periodic Anderson model complementally.

\section{Summary}
We have considered the extended periodic Anderson model by combining
dynamical mean-field theory with the non-crossing approximation.
Calculating the number of $f$-electrons at each site, 
we have discussed
how valence fluctuations are enhanced by the Coulomb interaction between
the conduction and $f$-electrons.
We have studied the stability of the valence transition with a jump singularity
in the particle number of $f$ electrons at finite temperatures complementally.
It is also interesting how the magnetic field induces
the metamagnetic transition~\cite{Sugibayashi},
which is now under consideration.

\section*{Acknowledgements} 
This work was partly supported by Japan Society for the Promotion of Science 
Grants-in-Aid for Scientific Research Grant Number 25800193 (A.K.) and 
the Global COE Program ``Nanoscience and Quantum Physics" from
the Ministry of Education, Culture, Sports, Science and
Technology (MEXT) of Japan.

\end{document}